\documentclass[ preprint, 5p, twocolumns,sort&compress]{elsarticle}

\usepackage{siunitx}
\usepackage{color}
\usepackage{amsmath}
\usepackage{lineno}

\title{Energy response of X-rays under high flux conditions using a thin APD for the energy range of 6--33\,keV}

\begin{document}

\begin{frontmatter}

\author[okayama]{T.~Masuda}
\ead{masuda@okayama-u.ac.jp}

\author[okayama]{T.~Hiraki}
\author[okayama]{H.~Kaino}
\author[kek]{S.~Kishimoto}
\author[okayama]{Y.~Miyamoto}
\author[okayama]{K.~Okai}
\author[okayama]{S.~Okubo}
\author[okayama]{R.~Ozaki}
\author[okayama]{N.~Sasao}
\author[okayama]{K.~Suzuki}
\author[okayama]{S.~Uetake}
\author[okayama]{A.~Yoshimi}
\author[okayama]{K.~Yoshimura}

\address[okayama]{Research Institute for Interdisciplinary Science, Okayama University, Okayama, Japan}
\address[kek]{High Energy Accelerator Research Organization, Tsukuba, Ibaraki, Japan}

\begin{abstract}
 This paper reports on the demonstration of a high-rate energy measurement technique using a thin depletion layer silicon avalanche photodiode (Si-APD). 
 A dedicated amplitude-to-time converter is developed to realize simultaneous energy and timing measurement in a high rate condition.
 The energy response of the system is systematically studied by using monochromatic X-ray beam with an incident energy ranging from 6 to \SI{33}{keV}. 
 The obtained energy spectra contain clear peaks and tail distributions. The peak fraction monotonously decreases as the incident photon energy increases. This phenomenon can be explained by considering the distribution of the energy deposit in silicon, which is investigated by using a Monte Carlo simulation.
\end{abstract}


\end{frontmatter}

%\linenumbers

\section{Introduction}
\label{sec:intro}
Silicon avalanche photodiodes (Si-APDs) are extensively used in a variety of experiments because of their single photon sensitivity, magnetic field insensitiveness, robustness, and fast time response\cite{7100938,OHNO2016410,NEILSON200968,0031-9155-55-7-003,1748-0221-10-03-P03008,4696575}. Although most of the applications use Si-APDs as highly sensitive visible or near-infrared light sensors, they are able to detect X-rays at a single photon level with good signal-to-noise ratio.

Si-APDs are used both with and without scintillators for X-ray detection.
In particular, direct X-ray detection without any scintillators is used for experiments that need fast time response for a single photon\cite{Kishimoto2010,Kataoka2010,Monteiro2018}. Si-APD without a scintillator can realize satisfactory time resolution of typically $\sim$ns in the full width at half maximum (FWHM), and higher than \SI{100}{ps} was also reported\cite{Baron2006}. A high count rate capability of more than $10^6$--$10^8$ counts per second (cps) can also be easily achieved using APDs. It is difficult to sustain such high counting rate by using scintillators.
The energy resolution of Si-APD itself is another advantageous characteristic. Si-APDs operated in the linear region present an energy resolution of approximately 10--20\% in FWHM. Operation involving both the fast time response and energy sensitivity is one of the advantages of Si-APD for direct X-ray detection. 

It is well known that the internal structure of Si-APD strongly affects the spectra of both time and energy.
The photon absorption position widely distributes along the thickness because the attenuation length of photons in silicon is of nearly the same order as that of the Si-APD thickness in the energy region of a few to tens of keV. The drift velocity of electrons in silicon is approximately \SI{100}{\micro m/ns} at \SI{10}{V/\micro m} and room temperature\cite{Jacoboni1977}, thereby limiting the time resolution. The thickness of the drift region therefore, should be less from the viewpoint of timing performance.
The energy response has a different dependence on the structure. The output pulse height of Si-APD irradiated by monochromatic X-rays generally has a low-energy tail component as well as a full energy peak. The presence of the tail component is potentially due to the photons absorbed not at the drift region but those at the avalanche region\cite{Yatsu2006}; thus, a thicker drift region is better for energy measurement. Many studies have focused on the energy response of various Si-APDs for X-rays\cite{Fernandes2007,CheeHingTan2011};however, because of the incompatibility between the energy and timing response, there are few reports on the energy response of the Si-APDs optimized for fast time response.

We recently developed a fast single X-ray photon detection system by using thin Si-APDs\cite{Masuda2017}. It was developed for the synchrotron-radiation based nuclear resonant scattering (NRS) measurement of thorium-229\cite{Yoshimi2018}. The aim of the experiment is population transfer from the nuclear ground state to the nuclear first excited state whose energy is extremely low\cite{PhysRevLett.98.142501,VonderWense2016}. The process that we use is nuclear resonant excitation to the second excited state followed by deexcitation to the first excited state. The NRS signals are detected to confirm the population transfer. The NRS measurement is carried out in the time domain as is usually done with general NRS measurements. The problem is the expected short lifetime and the narrow partial width for the resonant excitation of the objective state --- only $\sim$\SI{150}{ps} and $\sim$\SI{1}{neV}, respectively. To overcome this problem, the detection system needs quite fast time response and high rate tolerance. In addition, it needs to measure the energy of a single photon simultaneously to suppress the background due to the radioactivity of thorium-229 and its daughter nuclei.
The developed system had an overall time resolution of \SI{120}{ps} in FWHM and a short tail of the order of $10^{-9}$ at \SI{1}{ns} apart from the peak in the time spectrum. It could also measure both the energy and timing of a single photon simultaneously, even at a high rate of more than \SI{E6}{cps} for one Si-APD. 

 In the previous work, we focused mainly on the timing performance of the system by using actual NRS data. In this paper, however, we describe the method of both the timing and energy measurement in high count rate, and the obtained performance by using dedicated monochromatic X-ray beam whose energy is tuned from 6 to 33\,keV. A Monte Carlo simulation is also performed to explain the obtained energy spectra.

\section{Energy measurement method}\label{sec:atc}
 A technique commonly used to measure the energy is to integrate the charge of the detector signal by a gated integrator followed by processing by an analog-to-digital (AD) converter or a multichannel analyzer. AD conversion of shaped pulse height, which is proportional to the charge, is also used.
 For cases in which both the energy and timing need to be recorded, amplitude-to-time converters (ATCs) or charge-to-time converters followed by TDC are also widely used\cite{Fujita1998,Bespalko2009,Nishino2009,Liu2010,Parl2012}. This method can further simplify a method compared to a system built with both ADCs and TDCs. The pulse shape recording technique is impractical for such high rate measurement of around or above \SI{1E6}{cps} per channel owing to a large amount of data. The ATC method is a practical solution.

\begin{figure}[htbp]
\centering
\includegraphics[width=0.4\textwidth, bb=0 0 80 30, clip]{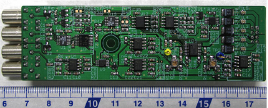}
\caption{\label{fig:atcpicture}(color online) Printed circuit board of one channel of ATC.}
\end{figure}

\begin{figure*}[htbp]
\centering
\includegraphics[width=0.9\textwidth, bb=0 0 830 350, clip]{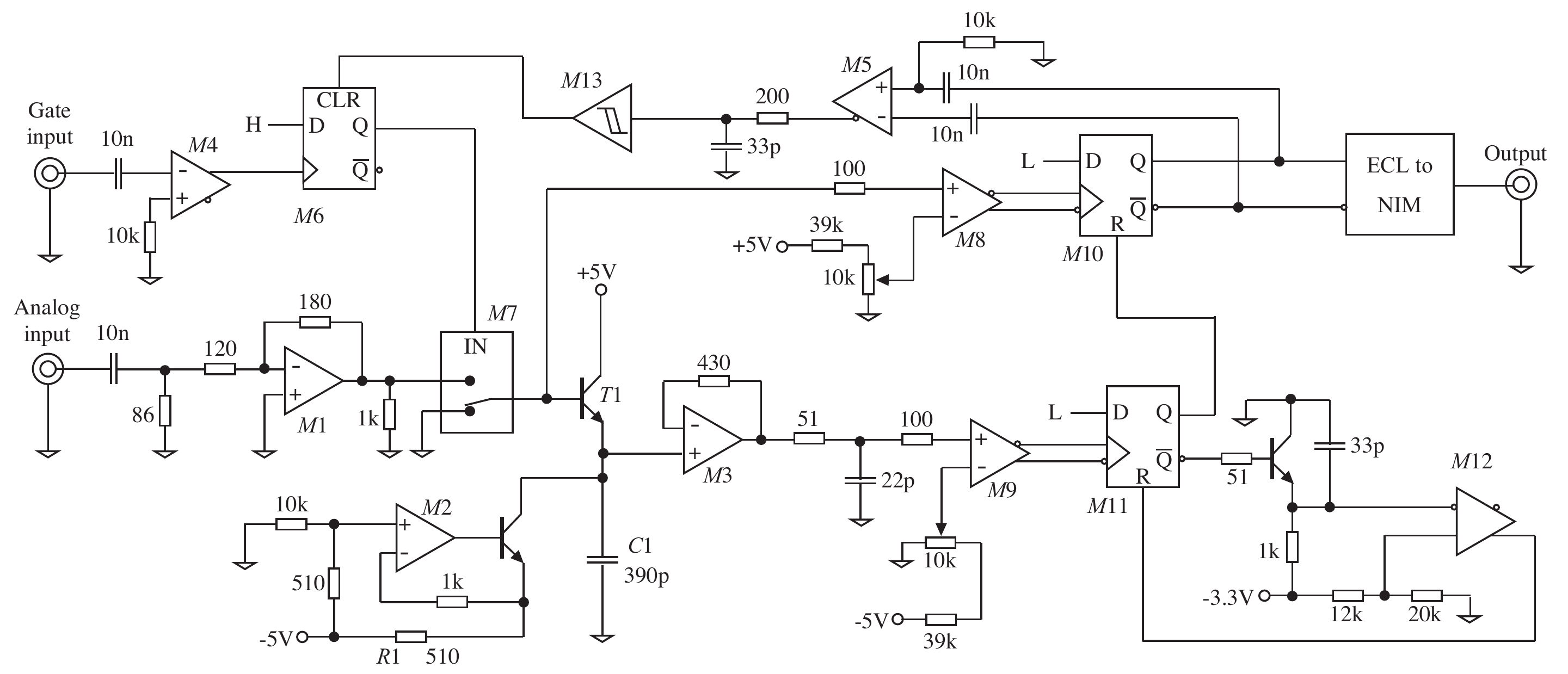}
\caption{\label{fig:atccircuit}Simplified ATC circuit diagram. The power supply lines, pull-up (down) resisters, bypass capacitors, electrostatic discharge protection components, and other less relevant components are omitted. \textit{M}1--3: operational amplifiers (AD8000), \textit{M}4--5: Logic converters (LTC6957), \textit{M}6: CMOS D-type Flip-Flop (TC74LCX74), \textit{M}7: Analog switch (MAX4644), \textit{M}8--9: ECL comparators (MAX9693), \textit{M}10-11: ECL D-type Flip-Flops (MC100LVEL51), \textit{M}12: ECL buffer (MC100LVEL16), \textit{M}13: Schmitt buffer (NL27WZ17), and \textit{T}1: NPN transistor (2SD2656). H and L represent logic high and low, respectively.}
\end{figure*}

\begin{figure}[htbp]
\centering
\includegraphics[width=.4\textwidth, bb=0 0 720 550, clip]{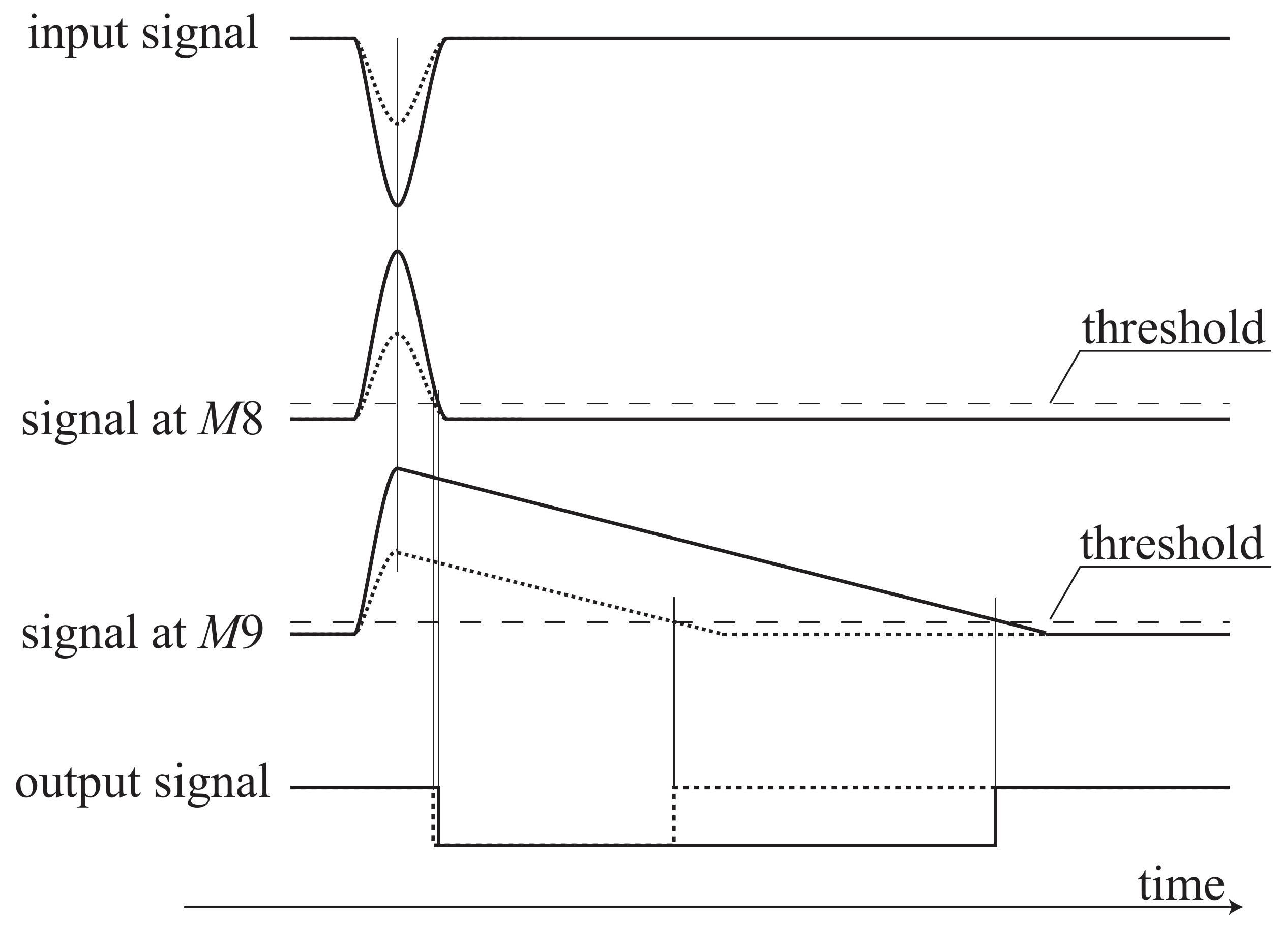}
\caption{\label{fig:sequence}Time sequence of the ATC operation demonstrating two cases for pulses with different amplitudes. The solid (dashed) lines represent a higher (lower) pulse height case.}
\end{figure}

 The developed ATC consists of a peak-hold circuit, a constant discharge circuit, and peripheral logic circuits.
 Figure~\ref{fig:atcpicture} shows an image of the printed circuit board of ATC for a channel.
 Figure~\ref{fig:atccircuit} shows the schematic diagram of the ATC. 
The converter operates as follows. 
The transistor \textit{T}1 is open in the initial position and the constant current flows from the +\SI{5}{V} source to the \SI{-5}{V} sink. A start signal and a negative analog pulse are input from the input connectors; subsequently, the analog switch \textit{M}7 changes the connection and the positive pulse converted by the operational amplifier \textit{M}1 goes to \textit{T}1.
During the increase in the positive pulse, \textit{T}1 is open and the capacitor \textit{C}1 is immediately charged by the current from the +\SI{5}{V} source. When the pulse is decreasing, \textit{C}1 maintains the voltage at its peak value and \textit{T}1 closes. Then, \textit{C}1 constantly discharges through the resistor \textit{R}1, and returns to the initial condition.
One of the comparators, \textit{M}8, toggles the output logic state to high when the end point of the positive pulse arrives at \textit{T}1, and the second comparator \textit{M}9 toggles the output logic state to low when the conversion is completed by the following logic circuit. The state of the analog switch \textit{M}7 is changed immediately after the analog pulse passes through it, and it is maintained during the conversion to ignore the next signal coming before the conversion is completed. The timing sequence of the ATC is shown in Fig.~\ref{fig:sequence}.

The ATC demonstrated a conversion time of \SI{10}{ns/keV}. This value can be reduced by decreasing the value of \textit{C}1 or \textit{R}1 by a certain degree. We selected \SI{390}{pF} as \textit{C}1 because we noted a deterioration of the energy response when \textit{C}1 was \SI{100}{pF}.

The energy spectra obtained in the actual NRS measurement at high count rate of more than \SI{E6}{cps} have been reported in our previous paper\cite{Masuda2017}.
This is potentially useful for other types of measurements as well as the $^{229}$Th NRS measurement, for example, depth selective M\"{o}ssbauer spectroscopy by electron detection proposed in \cite{Seto2017}.

\section{Energy response study}
 This section presents the energy responses of the Si-APD as a direct X-ray sensor. The absolute efficiency was measured at an energy of \SI{13}{keV}, which is our research object in the thorium-229 NRS experiment. The internal structure can be estimated by using the efficiency data. The incident energy dependence of the energy spectra was considered for values from 6 to \SI{33}{keV}.

\subsection{Setup}
 The experiments described in this section were performed at the BL-14A beam line of KEK Photon Factory (KEK-PF)\cite{doi:10.1063/1.1140728}. The incident X-ray beam from the storage ring was monochromatized by a Si(111) double-crystal monochromator. The typical relative bandwidth $\Delta E/E$ of the beam is $2\times 10^{-4}$. A Si(553) monochromator was used above \SI{24}{keV}. The beam energy was tuned from 6 to \SI{33}{keV}. Pin-holes located at the upper stream of the Si-APD defined the beam size. Several filters made of pure metals were used to moderate the beam intensity.
 The beam was injected directly into the Si-APD perpendicular to the surface.
 
The Si-APD (Hamamatsu Photonics S12053-05) whose sensitive area was \SI{0.5}{mm} in diameter, is a reach-through type Si-APD. The detector system was almost the same as that used in the previous work but the number of Si-APD was changed from six to seven, and an associated minor modification of a preamplifier circuit layout was carried out. One out of the seven Si-APDs was tested in this measurement because they are identical, including the circuit. The device was operated at room temperature. The applied reverse bias voltage was \SI{150}{V} and the nominal gain was 50.

The output signal was amplified by a preamplifier (Mini-circuits, RAM-8A+) whose gain was \SI{32}{dB}, and it was sent to the CFD and a pulse shaping circuit. The CFD determined the pulse timing within a deviation of \SI{\pm25}{ps}, and also provided a start signal to the following ATC described in Sect.~\ref{sec:atc}. Both the CFD and ATC output and an accelerator reference signal were digitized by a multi-stop TDC (FAST ComTec, MCS6) at a sampling rate of 100-ps. The recorded data were converted to timing and energy information for each signal in the offline analysis. The energy information was converted from the time difference between an ATC output pulse and the previous CFD output pulse.

\subsection{Efficiency at \SI{13}{keV}}\label{sec:efficiency}
 The absolute efficiency of the Si-APD was measured by limiting the beam size to \SI{0.4}{mm} in diameter by a pin-hole so that the sensitive area of the Si-APD covered the whole beam. The absolute photon flux was measured by using a NaI scintillator with a 150-\SI{}{\micro m}-thick beryllium window. Because the NaI scintillator is a low count rate detector, we placed additional zirconium filters in front of the NaI scintillator to reduce the counting rate to less than \SI{E4}{cps}. The transmittance of the filters was estimated by changing the filter combination. The scintillation lights were read by a photomultiplier tube and were digitized by a multi-channel analyzer.
 
\begin{figure}[htbp]
\centering
\includegraphics[width=.45\textwidth, bb=0 0 580 380, clip]{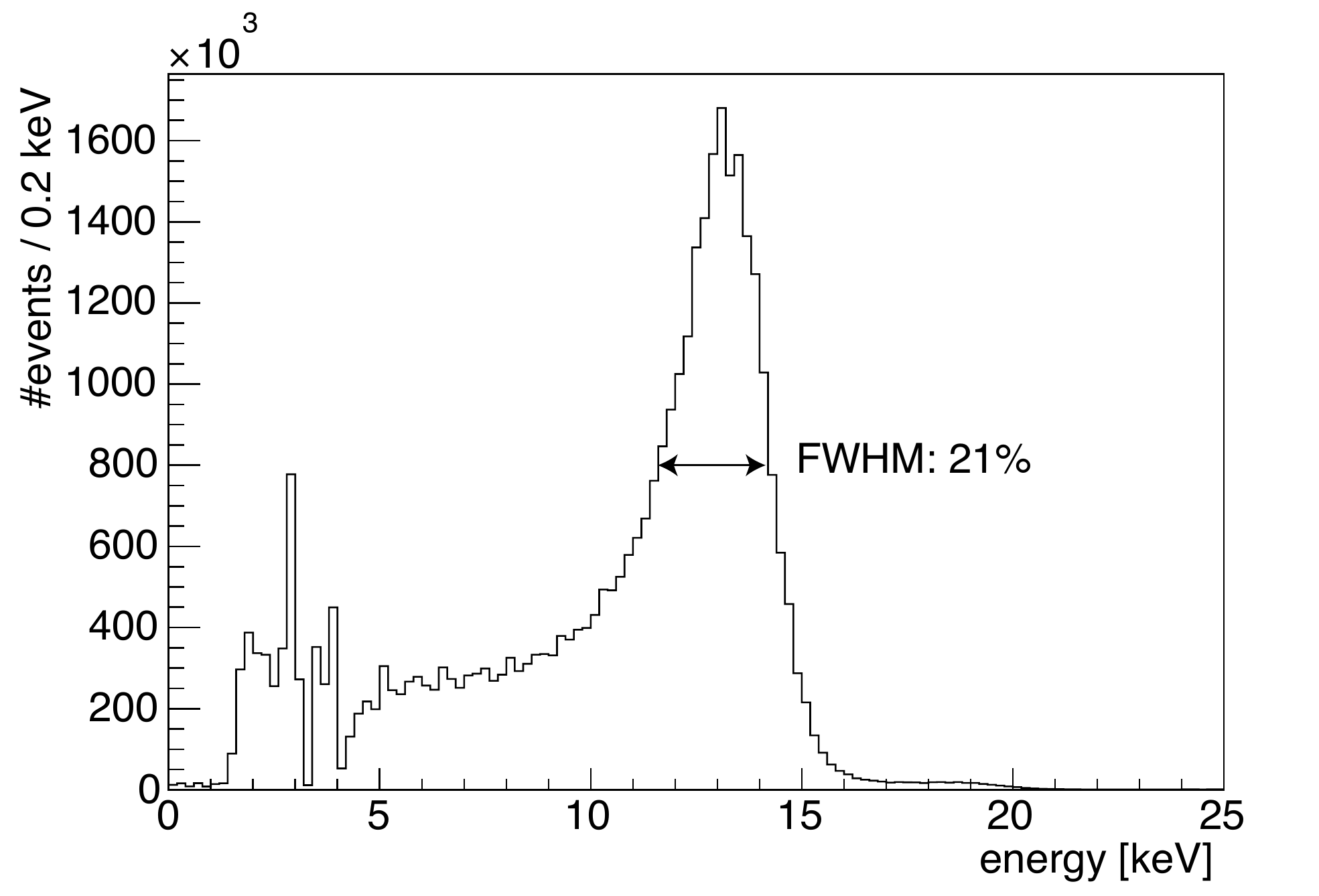}
\caption{\label{fig:pinhole04}Observed energy spectrum for \SI{13}{keV} X-rays.}
\end{figure}

 The energy spectrum measured by the Si-APD is shown in Fig.~\ref{fig:pinhole04}. The counting rate was \SI{3.7E5}{cps} and the measuring time was \SI{100}{s} for each spectrum. There was a clear peak at \SI{13}{keV}; the FWHM was 21\%. The low energy tail was also observed. The fluctuation at the 3--4\,keV region was caused by digital electric noise in the ATC circuit board.
 
 The absolute efficiency was measured by counting the count rate of the Si-APD. The efficiency of the whole region including the tail component was approximately 1.7\%, which corresponded to a silicon thickness of \SI{4.8}{\micro m}; this was calculated based on the efficiency and the linear attenuation coefficient of silicon\cite{xcom}. If energy selection is required, the efficiency in the full energy peak is also important. The efficiency within the $\pm 20\%$ region from the peak was 1.0$\pm$0.3\%, and it corresponded to a silicon thickness of \SI{2.9}{\micro m}.
 
\subsection{Energy spectra}
\begin{figure}[htbp]
\centering
\includegraphics[width=.45\textwidth, bb=0 0 550 370, clip]{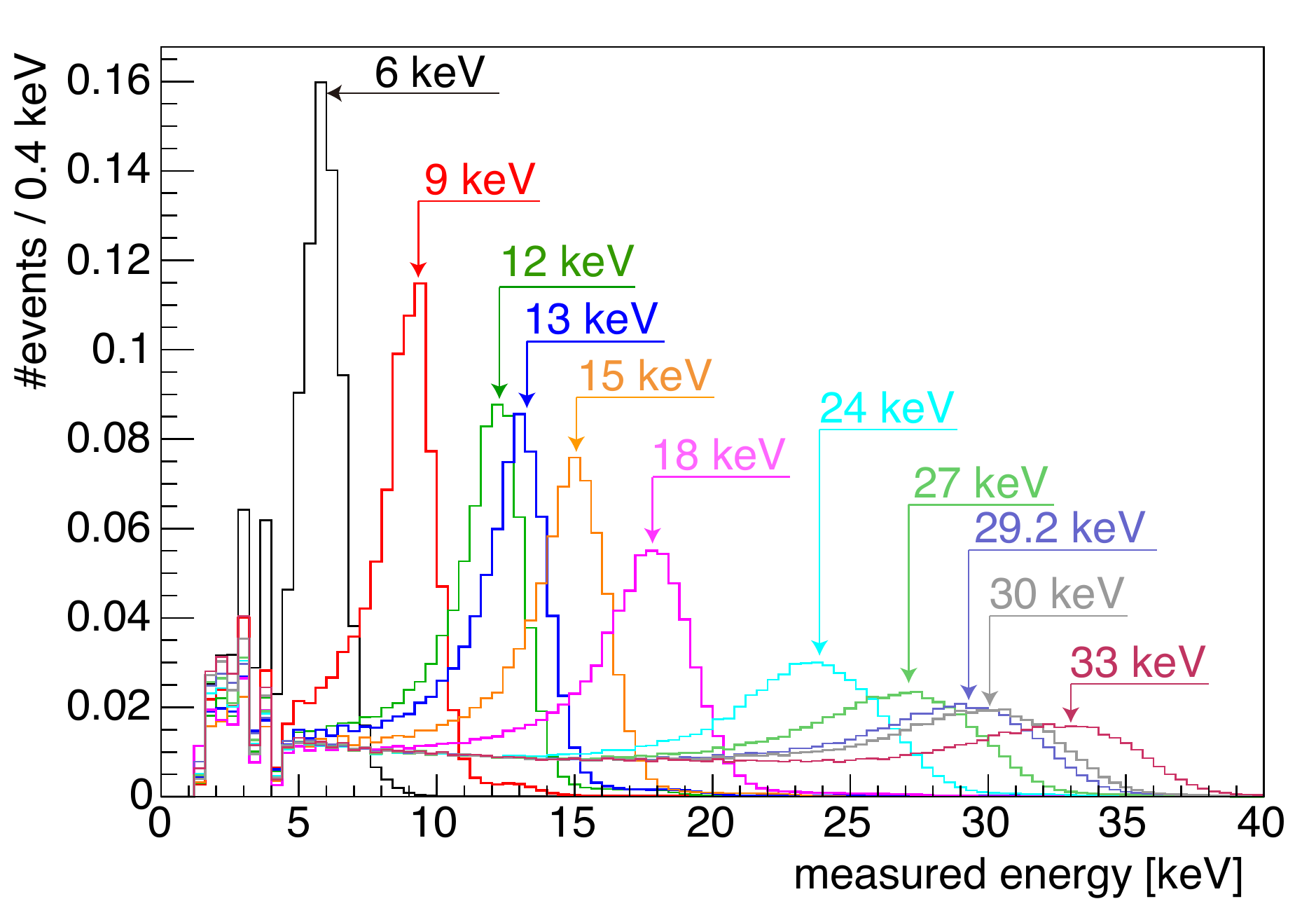}
\caption{\label{fig:collectedspectra}(color online) Energy spectra with linearity correction. All histograms are normalized so that their integration is 1. Each label represents the incident X-ray energy for each histogram.}
\end{figure}

 Energy spectra were obtained for various incident X-ray energies: 6.0, 9.0, 12.0, 13.0, 15.0, 18.0, 24.0, 27.0, 29.2, 30.0, and 33.0\,keV. The beam size was \SI{0.8}{mm} in diameter in these measurements and thus, the sensitive area was fully irradiated similar to an actual situation. Figure~\ref{fig:collectedspectra} shows the energy spectra. The peak-to-tail ratio decreases in a pronounced manner with increase in the incident energy. 
On the other hand, the relative widths which are the ratio of the peaks width and the peak values are almost constant, between 20--30\%. Despite this system being optimized for timing performance, it can still obtain energy information at this level.

\begin{figure}[htbp]
\centering
\includegraphics[width=.45\textwidth, bb=0 0 550 370, clip]{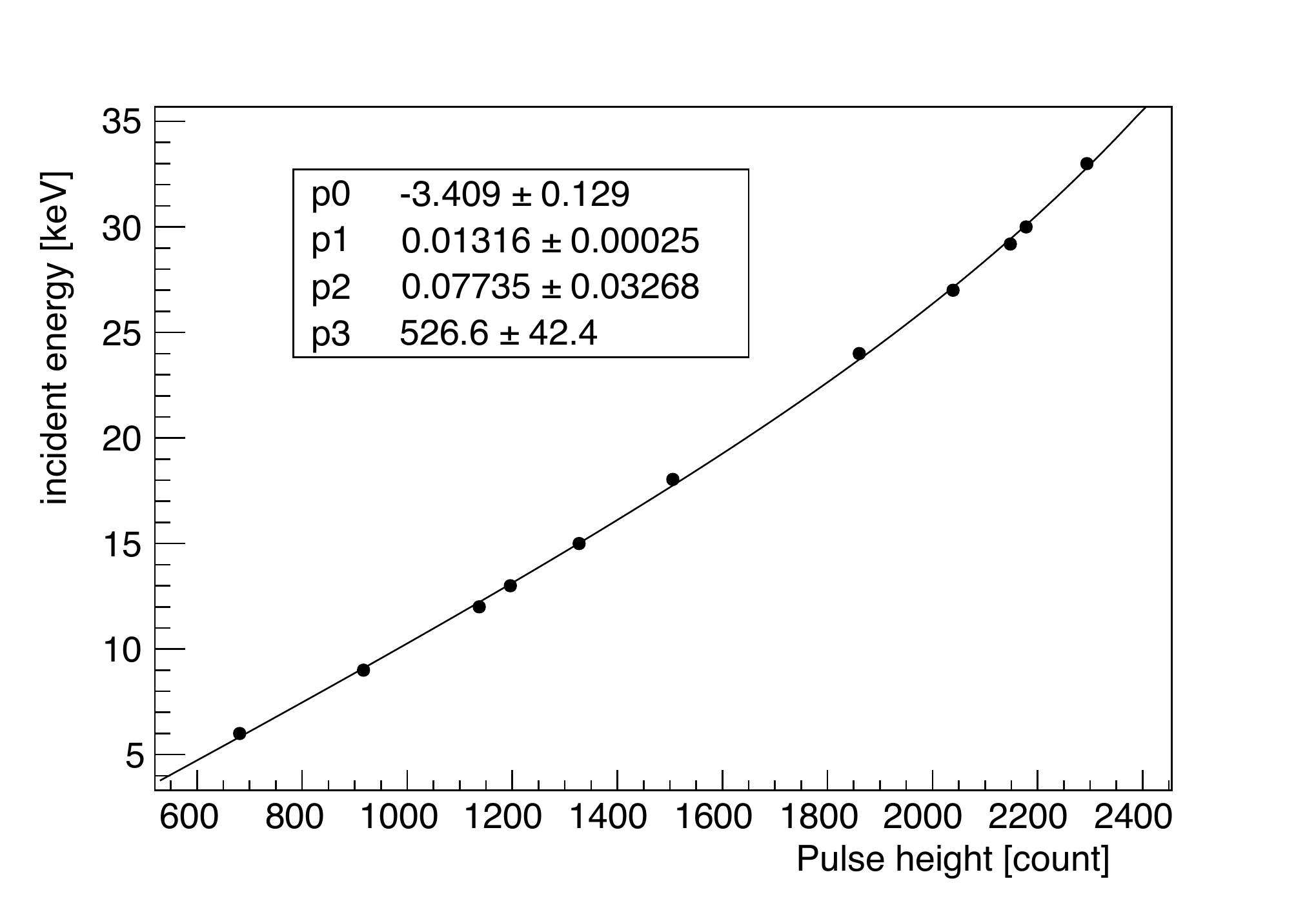}
\caption{\label{fig:linearity}Incident energy vs. pulse height. The horizontal axis represents the peak centers of the ATC spectra. One count corresponds to \SI{0.1}{ns}. The vertical axis indicates the energy of the incident X-ray beam. The solid points denote data for each incident energy. The curve shows the fit function described in the text, and the parameters in the box are the fit results.}
\end{figure}

We noted a slight nonlinearity in the energy spectra. Figure~\ref{fig:linearity} shows the relation between the measured values at the peaks and the actual incident X-ray energy. The values at the peaks were obtained by Gaussian fit at the peaks. We checked the preamplifier output pulse shapes for each incident energy and they did not demonstrate such degradation. This nonlinearity was therefore primarily owing to the shaper and the ATC saturation and was not caused by the Si-APD itself. 
Note that the nonlinearity of the spectra shown in Fig.~\ref{fig:collectedspectra} was already corrected. We used an empirical function of $p_0 + p_1 x + p_2 \exp (x/p_3)$ for the correction. The fit results are shown in Fig.~\ref{fig:linearity}.

\section{Discussion}

\subsection{Peak width}
\begin{figure}[htbp]
\centering
\includegraphics[width=.45\textwidth, bb=0 0 550 370, clip]{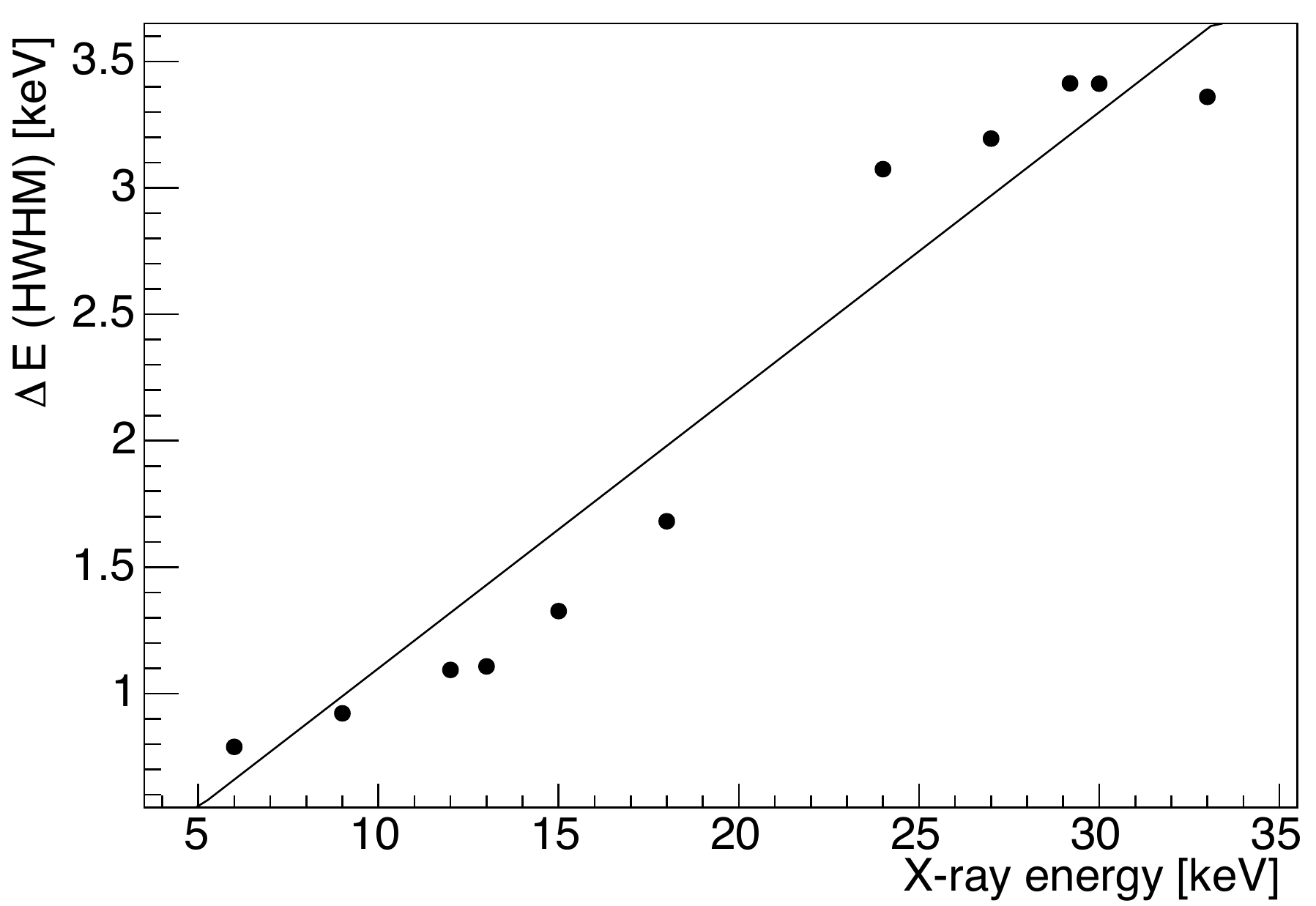}
\caption{\label{fig:hwhm}Peak width for each incident energy. The solid circles denote the data. The straight line represents the linear fit crossing the origin.}
\end{figure}

The peak width is defined as the half width at the half maximum (HWHM), which indicates the interval from the peak center to the half maximum point in the higher energy side. This is done to avoid the low-energy tail effect. The peak width for each incident energy is shown in Fig.~\ref{fig:hwhm}, and it exhibits a roughly linear behavior. The FWHM energy resolution, which is twice the peak width, is approximately 20--30\%.

The FWHM of a full energy peak $\Delta E$ of a Si-APD in X-ray detection can be written as follows\cite{Fernandes2007}:
\begin{align*}
\Delta E &= \\
 &  2.35 \left[ E \varepsilon (f+F-1) + \left( \frac{\sigma_{MU}}{M} \right)^2 E^2 + (\Delta E_\mathrm{n})^2 \right]^{1/2},
\end{align*}
where $E$ is the energy of the peak, $\varepsilon=$ \SI{3.65}{eV} is the unit energy generating one electron-hole pair in silicon, and $f$ is the Fano factor, which can be assumed to be $\sim 0.1$ for a silicon detector\cite{Lowe1997}. $F$ is the excess noise factor; it depends on the gain and it is of the order of one\cite{Yatsu2006}. The term $\sigma_{MU} / M$ refers to the gain deviation which is caused by non-uniformity of the avalanche gain and the system. $\Delta E_\mathrm{n}$ is the noise contribution. The first term in the square bracket, which is the statistical term, contributed only less than 5\% in our measurement and the noise term does not depend on the energy; thus, the observed energy resolution was dominated by the gain deviation.

\subsection{Peak fraction}
\begin{figure}[htbp]
\centering
\includegraphics[width=.45\textwidth, bb=0 0 550 420, clip]{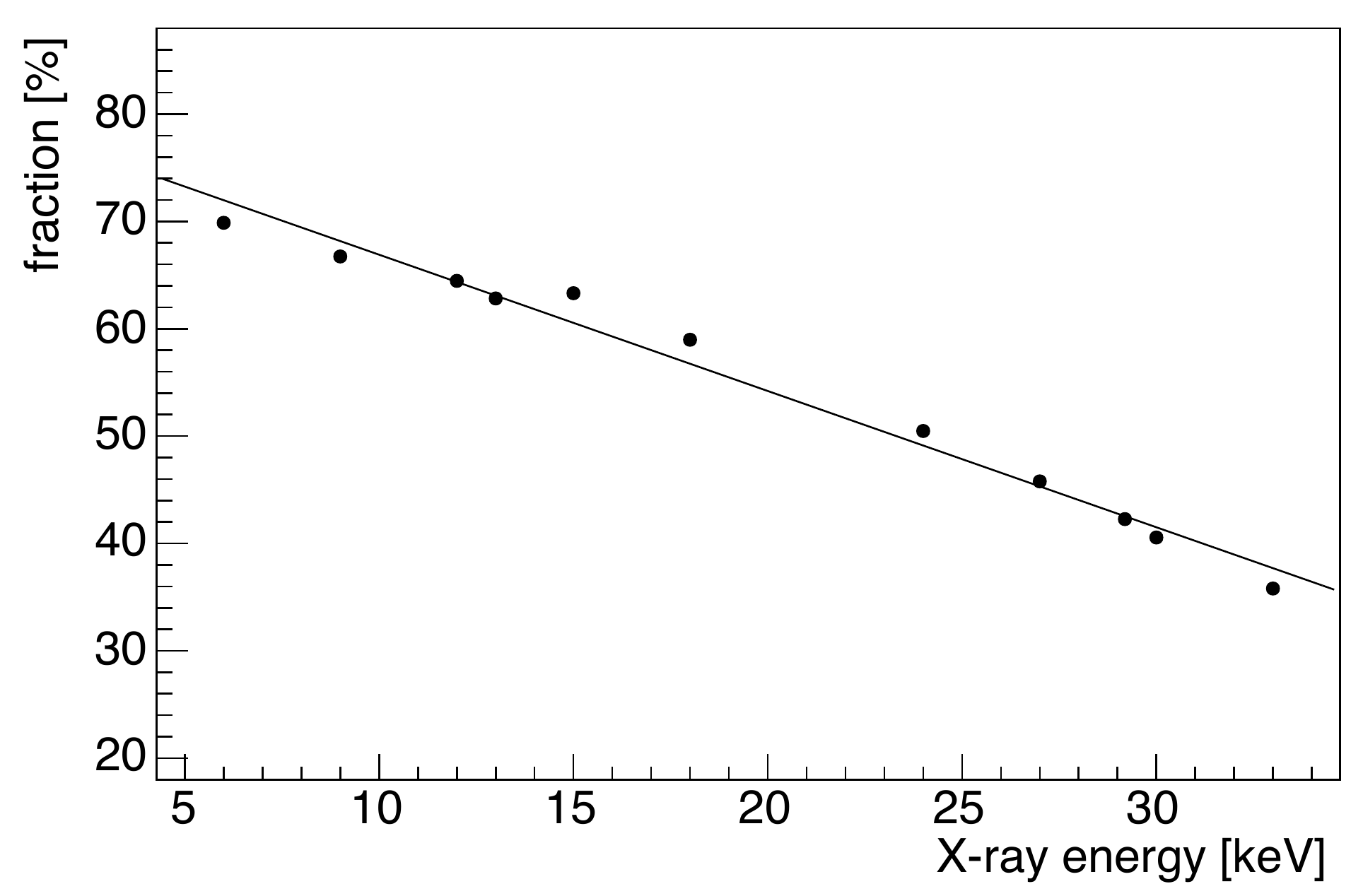}
\caption{\label{fig:peakfraction}Incident energy dependence of the peak fraction. The solid circles denote the data and the straight line represents the linear fit.}
\end{figure}

Figure~\ref{fig:peakfraction} shows the incident energy dependence of the peak fraction. Here, the peak fraction is defined as the number of events in the energy region within two times of the half width from the center to both sides divided by the number of events corresponding to more than \SI{2}{keV}. A monotonically decreasing trend is noted. This behavior cannot be simply explained on the basis of the X-ray attenuation length and the position dependence of the avalanche gain along the thickness. The absorbed positions of irradiated photons uniformly distribute in the depletion layer because the thickness of the depletion layer is less than \SI{5}{\micro m} and the attenuation length in silicon is, for example, \SI{280}{\micro m} for \SI{13}{keV} photons\cite{xcom}. That means the peak fraction would be constant in the high energy region.

\begin{figure}[htbp]
\centering
\includegraphics[width=.35\textwidth, bb=0 0 650 530, clip]{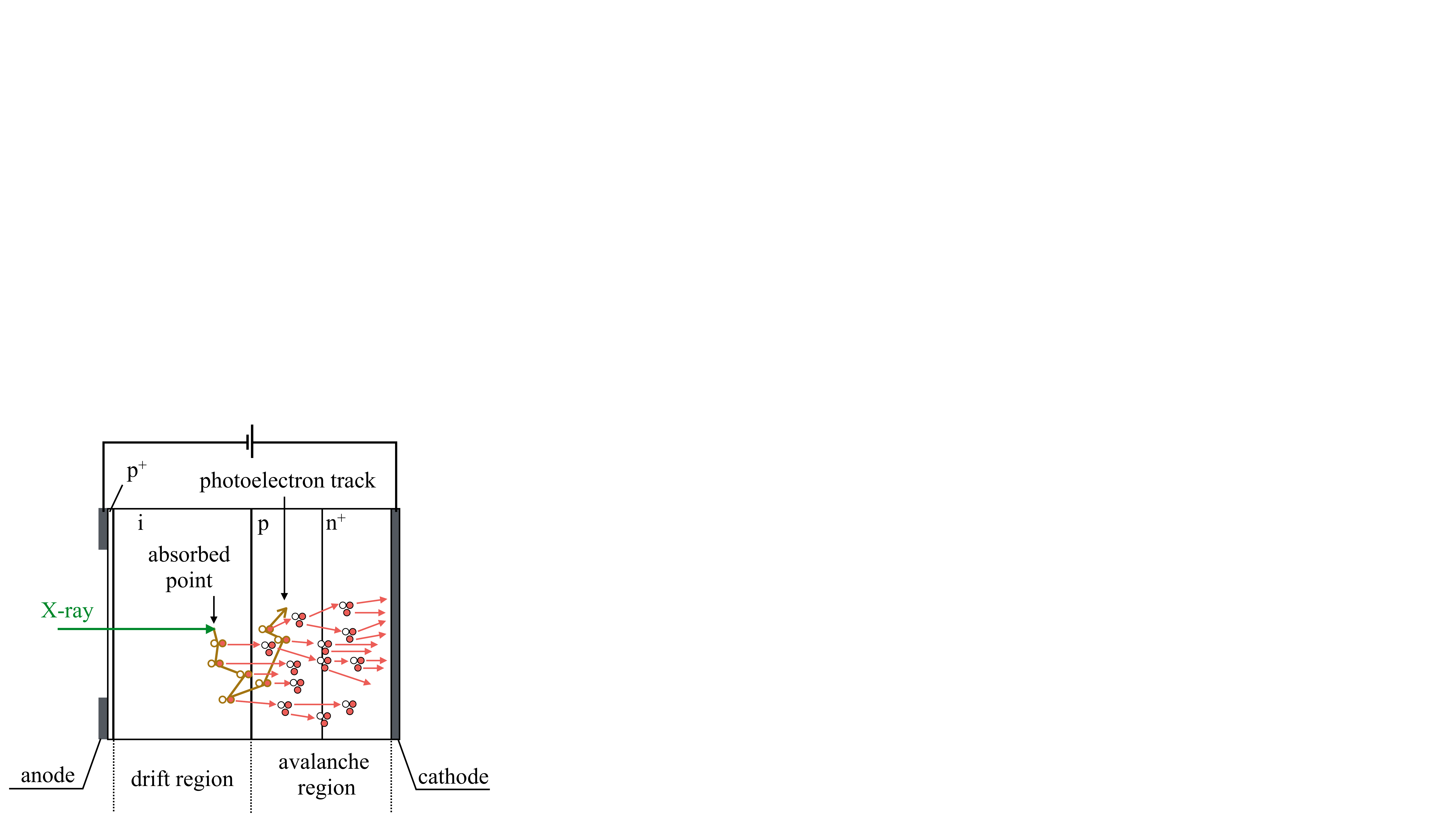}
\caption{\label{fig:manga}(color online) Schematic of the microscopic view of the reach-through type Si-APD operation for an X-ray photon. The X-ray photon comes in from the left side and is absorbed in the Si-APD. The photoelectron that has sufficient kinetic energy moves inside randomly owing to the multiple scattering process. The track of the photoelectron is represented by the dark yellow zig-zag line. Many electron-hole pairs are created along the track. The red (white) circles represent electrons (holes). The electrons are transferred to the cathode side as indicated by the red arrows and multiplied due to the high electric field in the avalanche region.}
\end{figure}

This tendency may be explained by considering the travel of initial photoelectrons in silicon, as shown in Fig.~\ref{fig:manga}. The photoelectron originating at the photoelectric absorption point has energy equal to the energy difference between the incident photon energy and the binding energy of the K-shell electron of silicon. This photoelectron moves in the silicon and creates many electron-hole pairs along its track. In the usual case in which the energy is low and a Si-APD is thick, there is no need to consider the spread due to the photoelectron travel;
 however, the spread becomes important in case the thickness of the Si-APD is low and the photon energy is high such that the photoelectron range is comparable with the Si-APD thickness. Photoelectrons that originate even in the drift region can move to the avalanche region and create electron-hole pairs; then, they contribute not to the full energy peak but to the tail component. The probability of occurrence of either depends on the range of electrons in silicon.

 A Monte Carlo simulation was performed to demonstrate the phenomenon. The simulation was based on the Geant4 (version 10.2) package\cite{Agostinelli2003,Allison2006,Allison2016} and its MuElec extension\cite{Valentin2012}. The extension can simulate electrons in silicon down to \SI{16.7}{eV} and the validated energy range is \SI{50}{eV}--\SI{50}{keV}. 

The simulation procedure is as follows: An absorption position along the depth was randomly generated for a photon based on the attenuation length at its incident energy. The energy deposit distribution along the photoelectron tracks was simulated by using the MuElec model. The output signal was generated by integrating the local energy deposits multiplied by the avalanche gain at the corresponding local points. Finally, the output signal was smeared by a gaussian distribution which corresponds to the gain deviation of $\sigma_{MU} / M$. We accumulated $10^5$ events for one energy set by repeating the procedure.
 
 The avalanche gain along the thickness $g(z)$ was simply assumed as the complementary error function, i.e., $g(z) \propto \mathrm{erfc} [(z-t)/d]$. Here, $z$ is the depth from the irradiated surface, $t$ corresponds to the depth at the center of the avalanche region, and $d$ corresponds to the width of the avalanche region. We assumed the parameters as $t = 3.95\,$\SI{}{\micro m}, $d=0.85$\,\SI{}{\micro m}, and $\sigma_{MU} / M=0.07$ to reproduce the absolute efficiency and the peak width at \SI{13}{keV} described in Sect.~\ref{sec:efficiency}.
 
\begin{figure}[htbp]
\centering
\includegraphics[width=.45\textwidth, bb=0 0 540 720, clip]{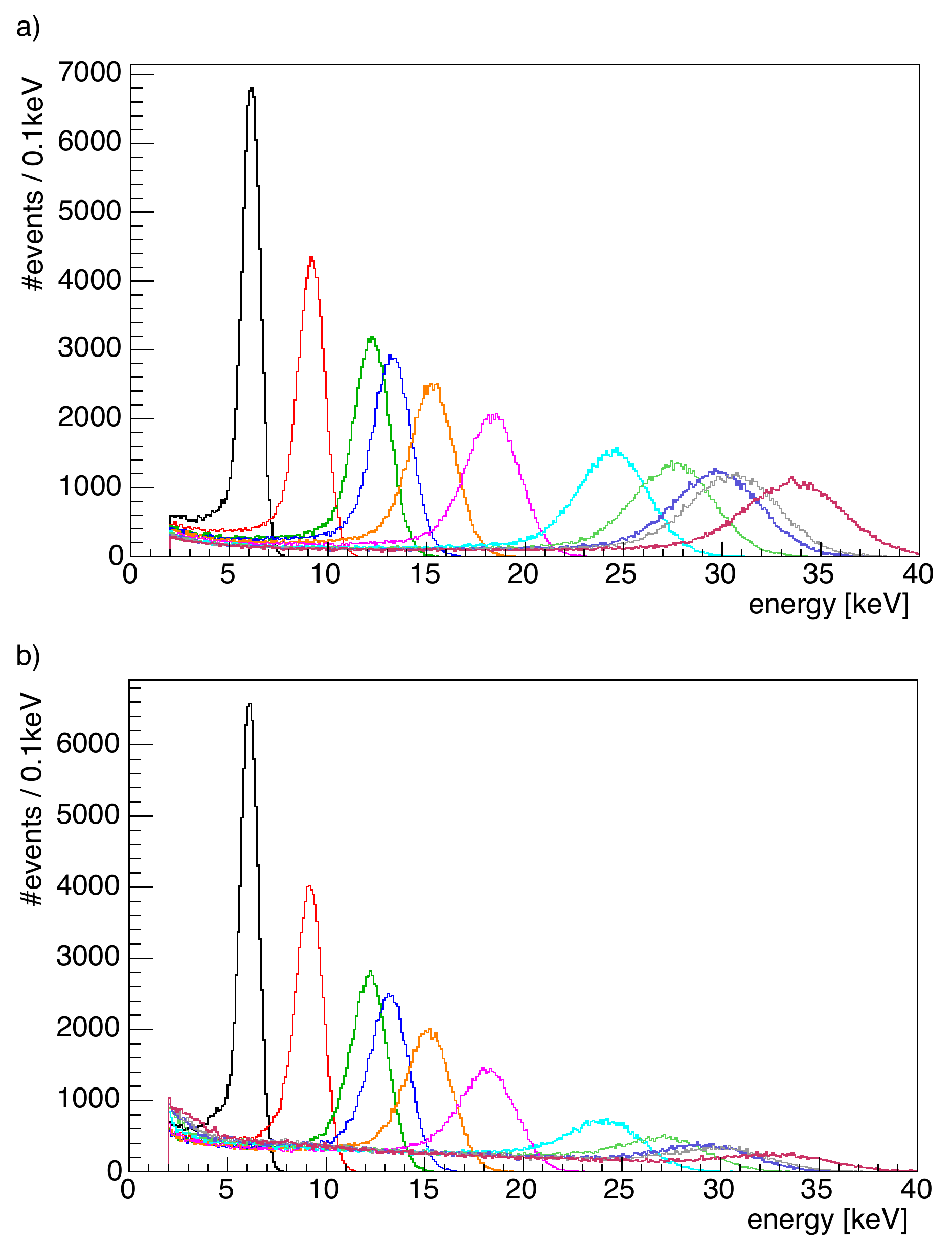}
\caption{\label{fig:spectrasim}(color online) Simulated energy spectra. a) Energy deposit spread not included. b) Energy deposit included. The color of each histogram is the same as that in Fig.~\ref{fig:collectedspectra}.}
\end{figure}

\begin{figure}[htbp]
\centering
\includegraphics[width=.45\textwidth, bb=0 0 550 420, clip]{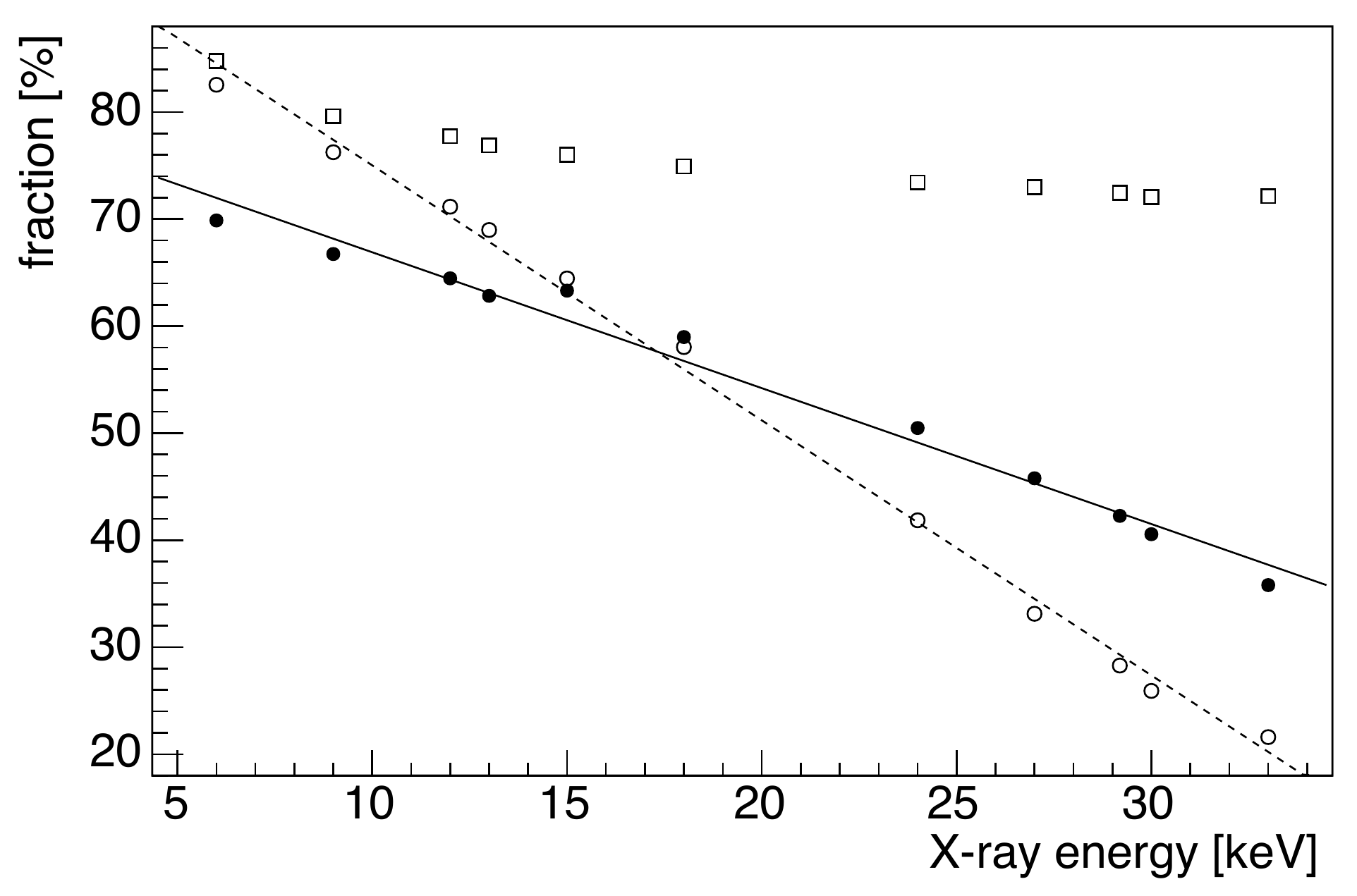}
\caption{\label{fig:peakfractionsim}Simulated peak fractions. The open circles correspond to the simulation including the energy deposit distribution in silicon and the open boxes correspond to the one without it. The dashed line represents the linear fit to the open circles. The solid circles and the solid lines are respectively the experimental data and the linear fit, as in Fig.~\ref{fig:peakfraction}. They are drawn for comparison.}
\end{figure}

 Figure~\ref{fig:spectrasim} shows the simulated energy spectra. The upper figure shows the spectra taking into account only the position dependence of the gain, the photon absorption position distribution, and the gain deviation. The spectra shown in the lower figure include the energy deposit distribution due to the photoelectron travel. A higher incident energy corresponds to a larger difference in the spectra. The peak fraction decreases owing to the energy deposit distribution, as shown in Fig.~\ref{fig:peakfractionsim}. This simulation can explain the rapid decrease in the peak fraction.
Because the internal gain distribution was simply assumed as to be a one-dimensional complementary error function in this simulation, knowledge of the three-dimensional internal structure may be useful to reduce the quantitative disagreement between the data and the simulation in the full energy region. In addition, it is possible that the imperfection of the simulation model might be the issue.
 
 The Si-APD in this study is thinner than those used in previous studies for energy measurement; therefore, the energy deposit distribution is relatively important for the consideration of the energy response especially for higher energy. The standard deviation of the energy deposit distribution is only \SI{0.09}{\micro m} along the depth in our simulation for the incident energy of \SI{6}{keV}; however, the corresponding value is \SI{2.67}{\micro m} for the incident energy of \SI{30}{keV}, which is comparable with the thickness of the depletion layer and thus cannot be ignored.
 
\section{Conclusion}
 We developed an X-ray photon counting system that equips thin Si-APDs. The system can accumulate both timing and energy of single X-ray photon simultaneously in high rate conditions. We developed a dedicated fast ATC that demonstrates a conversion time of only \SI{10}{ns/keV} to realize this requirement.
 Although the system was developed to maximize the timing performance, it exhibits reasonably good energy resolution using the ATC and a commercial TDC. 
 The absolute efficiency for \SI{13}{keV} photons was 1--2\%. 
 The FWHM energy resolution was 20--30\% in the energy region from 6 to \SI{33}{keV}. The peak fraction shows monotonous decreasing with respect to the incident photon energy. This phenomenon can be explained by considering the energy deposit distribution in silicon owing to the travel of the photoelectrons; this was observed by using a Monte Carlo simulation based on the Geant4 MuElec process. 

This work demonstrates the high rate energy measurement and the availability of the energy information with a Si-APD optimized for fast time response. It is seen that the energy deposit spread in silicon due to the travel of photoelectrons as well as the absorption position distribution of irradiated photons is important to understand the energy spectra obtained with such thin devices.

%\acknowledgments
\section*{Acknowledgments}
We thank M.~Seto and S.~Kitao for providing MCA6.
This work was performed under the approval of the Photon Factory Program Advisory Committee (Proposal No.~2017G085).
T.H. is supported by Grant-in-Aid for Japan Society for the Promotion of Science (JSPS) Research Fellow.
This work was supported by JSPS KAKENHI Grant Numbers JP15H03661, JP17K14291, JP18H01230, and JP18H04353. A.Y. acknowledges the MATSUO foundation.

\bibliographystyle{JHEP}
\bibliography{APD_EnergyResponse}
\end{document}